\documentclass[aps,pra,twocolumn,showpacs]{revtex4}
\usepackage{graphicx}
\usepackage{dcolumn}
\usepackage{bm}
\usepackage{color}
\usepackage{amssymb}
\usepackage{latexsym}
\usepackage{amsmath}

\begin{document}

\title{Thermodynamic properties of rotating trapped ideal Bose gases}

\author{Yushan Li${^{1,2}}$}

\author{Qiang Gu$^{1}$}
\address{1 Department of Physics, University of Science and Technology Beijing, Beijing 100083, China\\
2 Department of Physics, Heze University, Heze 274015, China}

\begin{abstract}
Ultracold atomic gases can be spined up either by confining them in
rotating frame, or by introducing ``synthetic" magnetic field. In
this paper, thermodynamics of rotating ideal Bose gases are
investigated within truncated-summation approach which keeps to take
into account the discrete nature of energy levels, rather than to
approximate the summation over single-particle energy levels by an
integral as does in semi-classical approximation. Our results show
that Bose gases in rotating frame exhibit much stronger dependence
on rotation frequency than those in ``synthetic" magnetic field.
Consequently, BEC can be more easily suppressed in rotating frame
than in ``synthetic" magnetic field.

\end{abstract}

\maketitle

\section {Introduction}

In the past few years, rotating Bose gases have been intensively
studied in the context of trapped ultracold atomic gases
\cite{Bloch2008,Fetter2009}. One may expect that rotation suppresses
Bose-Einstein condensation (BEC). Like superconductors in magnetic
field, vortices may be formed inside the condensate when it is
rotated and the Abrikosov vortex lattice can even be observed
\cite{Abo-Shaeer2001,Bretin2004,Schweikhard2004}. More strikingly,
rotating condensates in the fast-rotation limit are expected to
exhibit novel quantum phases analogous to the quantum Hall state of
electrons, if interactions between atoms are taken into account
\cite{Ho2001,Fischer2003,Watanabe2004,Wilkin2000,Cooper2001}.

Earlier experiments to spin up neutral atomic gases are to confine
them in a rotating frame \cite{Madison2000,Haljan2001,Hodby2002}.
More recently, the so-called ``synthetic" magnetic field approach is
developed \cite{Lin2009a,Lin2009b,Lin2011}, which creates an
effective gauge potential $\mathbf{A}$ for neutral atoms by means of
optical field \cite{Higbie2002,Jaksch2003,Juzeliunas2004}, instead
of rotating the frame. In this case, atoms behave like charged
particles in magnetic field. The ``synthetic" magnetic field has
already been realized in experiments and can be produced
significantly large, which makes it hopefully to achieve the
fast-rotating limit \cite{Lin2009a,Lin2009b,Lin2011}.

Although most research work on rotating Bose gases, both
experimental and theoretical, focuses on the ground state properties
or behaviors of rotating condensates, thermodynamic properties have
also stimulated a number of interests
\cite{Stringari1999,Kling2007,Pelster2010,Fan2011,Hassan2011,EI-Badry2013}.
The emphasis is mainly laid on Bose gases in the rotating frame. It
is reported that the critical temperature decreases with the
rotation frequency apparently
\cite{Stringari1999,Kling2007,Fan2011}. Recent attention was also
paid into the gases in ``synthetic" magnetic field
\cite{Fan2011,EI-Badry2013}.

The theoretical treatment in above works is mainly based on the
semi-classical approximation (SCA), in which the energy spectrum is
treated as a continuum and thus the summation over the discrete
single-particle energy levels is converted into phase-space
integrals. The SCA approach offers qualitatively accurate
descriptions of thermodynamics of Bose gases without rotation
through an extremely simple and efficient way \cite{Bagnato1987}.
Moreover, it is an effective approach to deal with Bose gases in the
rotating frame \cite{Stringari1999,Kling2007,Fan2011,Hassan2011}.
However, problem arises when the SCA approach is applied to ideal
Bose gases in ``synthetic" magnetic field \cite{Fan2011}. For
example, the BEC temperature in this case has no relation with the
magnetic field. In the meanwhile, the Landau diamagnetism also keeps
unchanging at all temperatures \cite{Fan2011}. These unphysical
results urge us to seek a more efficient treatment beyond the SCA
approach.

The direct way to go beyond the SCA is to reconsider the discrete
nature of the single-particle energy levels. Therefore, we compute
the thermodynamic quantities of rotating trapped ideal Bose gases by
performing the summation over discrete energy levels numerically by
truncating the summation at certain order. This truncated-summation
approach (TSA) produces reliable results and overcomes the
disadvantage of the SCA when applying to ideal Bose gases in
``synthetic" magnetic field.

This paper is organized as follows. Section 2 outlines main results
on the basis of the semi-classical approximation and discusses briefly
the problem that causes. We present numerical results based on the
truncated-summation approach in Section 3 and then conclude in the
last Section.

\section{The semi-classical approximation results}

\subsection{Trapped ideal Bose gases in the rotating frame}

The single-particle Hamiltonian describing rotating neutral bosons
with mass $M$ trapped in harmonic potential can be written as
\begin{align}
\hat{H}=&\frac{\mathbf{p}^2}{2M}+\frac{1}{2}M\omega^2_0\left(x^2+y^2\right)+\frac{1}{2}M\omega_z^2z^2-\Omega{\hat{L}_z}\nonumber\\
  =&\frac{\left(\mathbf{p}-\mathbf{A}\right)^2}{2M}+\frac{1}{2}M(\omega_0^2-\Omega^2)\left(x^2+y^2\right)+\frac{1}{2}M\omega_z^2z^2,
\label{eq:e1}
\end{align}
where $\mathbf{p}$ is the momentum operator, $\hat{L}_z$ is the
operator of the orbital angular momentum, $\omega_0$ and $\omega_z$
denote the transverse and axial frequencies of harmonic potential,
$\Omega$ is the rotation frequency around the z axis. The vector
potential in the symmetric gauge is
$\mathbf{A}=(\mathbf{B}\times\mathbf{r})/2=M(\mathbf{\Omega}\times\mathbf{r})$,
and $\mathbf{B}=2M\Omega\vec{e}_z$ is the effective ``magnetic"
field.

The eigen-energy $\varepsilon_{nlm}$ is given by
\begin{align}
\varepsilon_{nlm}=(n+\frac{1}{2}){\hbar\omega_z}+(2l+\left|{m}\right|+1){\hbar\omega_0}-m{\hbar\Omega}.
\end{align}
For simplicity, we can adopt a dimensionless treatment to
eigen-values $\varepsilon_{nlm}$ in units of $\hbar\omega_z$. One
has
\begin{align}
\bar{\varepsilon}_{nlm}=n+\frac{1}{2}+(2l+\left|{m}\right|+1)\sigma-m\sigma\xi,
\label{eq:e2}
\end{align}
where $\bar{\varepsilon}_{nlm}=\varepsilon_{nlm}/({\hbar\omega_z})$,
with the quantum numbers $n=0,1,2,\cdots$, $l=0,1,2,\cdots$, and
$m=0,\pm1,\pm2,\cdots$. $\sigma=\omega_0/{\omega_z}$ and
$\xi=\Omega/{\omega_0}$ denote the aspect ratio of harmonic trap and
the dimensionless rotation frequency, respectively. The ground state
energy of one particle is given by $\bar{\varepsilon}_0 =
\bar{\varepsilon}_{000}=1/2+\sigma$.

The dimensionless thermodynamic potential is expressed as
$\bar{\Omega}=\bar{\Omega}_T+\bar{\Omega}_0$. Thermodynamic
potential for thermal particles $\bar{\Omega}_T$ reads
\begin{align}
\bar{\Omega}_T&=\bar{T}\sum_{nlm}\ln\left[1-\exp\left(\frac{\bar{\mu}-\bar{\varepsilon}_{nlm}}{\bar{T}}\right)\right]\nonumber\\
  &\simeq\bar{T}\int_0^{\infty}dn\int_0^{\infty}dl\int_{-\infty}^{\infty}dm
    \ln\left[1-\exp\left(\frac{\bar{\mu}-\bar{\varepsilon}_{nlm}}{\bar{T}}\right)\right]\nonumber\\
  &=-\frac{\bar{T}^4}{\sigma^2(1-\xi^2)}g_4\left(\frac{\bar{\varepsilon}_{0}-\bar{\mu}}{\bar{T}}\right),
\label{eq:e3}
\end{align}
where the summation over energy levels is replaced by direct
integrals over all the quantum numbers. This is the so-called
semi-classical approximation. Here $\bar{T}=k_BT/(\hbar\omega_z)$ is
the dimensionless temperature, $\bar{\mu}=\mu/(\hbar\omega_z)$ is
chemical potential, and $g_\gamma{(z)}$ is the poly-logarithm
function which obeys the relation $\partial g_{\gamma}(z)/{\partial
x}=-g_{\gamma-1}(z){\partial z}/{\partial x}$. For condensed bosons,
$\bar{\Omega}_0$ is described by
\begin{align}
\bar{\Omega}_0=\frac{1}{\hbar\omega_z}\int d^3r
  \left\{{\frac{\left|\mathbf{D}\bar{\psi}\right|^2}{2M}}+V(r)\left|\bar{\psi}\right|^2-\mu\left|\bar{\psi}\right|^2\right\},
\label{eq:e4}
\end{align}
with $\mathbf{D}\bar{\psi}=\hbar\nabla\bar{\psi}+i\mathbf{A}\bar{\psi}$. $\bar{\psi}$ is a background field and $V(r)$ represents trapping potential.

All thermodynamic quantities can be obtained from the thermodynamic
potentials. The BEC temperature is given by
\begin{align}
\bar{T}_C=\left(\frac{\omega^2_0-\Omega^2}{\omega^2_zg_3(0)}N\right)^\frac{1}{3}=\left[\frac{\sigma^2(1-\xi^2)}{g_3(0)}N\right]^\frac{1}{3}.
\label{eq:e6}
\end{align}
Note that BEC temperature of Bose gases in the rotating frame
decreases with rotation frequency increasing. When the rotation
frequency $\Omega$ approaches to the trap frequency $\omega_0$
(called the fast-rotating limit), the BEC does no longer take place
\cite{Stringari1999,Kling2007,Fan2011}.

The condensate fraction is written as
\begin{align}
\frac{N_0}{N}=1-\left(\frac{\bar{T}}{\bar{T}_{C}}\right)^3.
\label{eq:e7}
\end{align}
This is formally identical with the expression of non-rotating
harmonically trapped Bose gases in three dimensional space. The
impact of rotating effect on the condensate fraction mainly comes
from Eq.~(\ref{eq:e6}).

If looking upon it as some effective magnetic field, $\xi$ may
induce ``magnetization", which reads
\begin{align}
\bar{M}=-\frac{\partial\bar{\Omega}}{\partial\xi}=-N\xi.
 \label{eq:e8}
\end{align}
The value of magnetization is negative, reflecting that Bose gases
in the rotating frame exhibit the Laudau diamagnetism. We point out
that the magnetization keeps invariant at all temperatures
\cite{Fan2011}, conflicting with our established intuition which the
magnetization varies with the temperature and diamagnetism is
stronger at lower temperatures.

\subsection{Trapped ideal Bose gases in ``synthetic" magnetic field}

The effective Hamiltonian for a trapped atom in ``synthetic"
magnetic field can be expressed as follows,
\begin{align}
  \hat{H}=\frac{\left(\mathbf{p}-\mathbf{A}\right)^2}{2M}+\frac{1}{2}M\omega_0^2\left(x^2+y^2\right)+\frac{1}{2}M\omega_z^2z^2,
\label{eq:e9}
\end{align}
which looks like that of charged particles in the magnetic field
with $\mathbf{A}$ representing ``synthetic" gauge potential. And the
energy levels are given by
\begin{align}
  \bar{\varepsilon}_{nlm}=n+\frac{1}{2}+(2l+\left|{m}\right|+1)\sqrt{\sigma^2+\bar{B}^2}-m\bar{B},
\label{eq:e10}
\end{align}
where $\sigma=\omega_0/{\omega_z}$, $\bar{B}={\omega_L}/{\omega_z}$, and $\omega_L$ denotes the Larmor frequency.

We can directly present the analytical expression of BEC temperature
within SCA framework according to Ref. \cite{Fan2011},
\begin{align}
\bar{T}_C=\left(\frac{\sigma^2N}{g_3(0)}\right)^{\frac{1}{3}}.
\label{eq:e11}
\end{align}
Obviously, BEC temperature is irrelevant to ``synthetic" magnetic
field, only determined by the frequency of harmonic potential and
the particle number.

Diamagnetization in ``synthetic" magnetic field $\bar{B}$ can be
calculated similarly to the rotating frame case,
\begin{align}
\bar{M}=-\frac{\partial\bar{\Omega}}{\partial\bar{B}}=-\frac{N\bar{B}}{\sqrt{\sigma^2+\bar{B}^2}}.
\label{eq:e12}
\end{align}
It vanishes as $\bar{B}\rightarrow0$, but seems to have no relation
with the temperatures. Therefore, a correction to thermodynamics of
the SCA is needed.

\subsection{Problem with the semi-classical approximation}
\label{sec1}

As already studied
\cite{Stringari1999,Kling2007,Fan2011,Hassan2011}, the SCA is very
useful in calculation the thermodynamics of Bose gases without
rotation and those in the rotating frame. It also provides a good
description of thermodynamic quantities such as the BEC temperature,
condensate fraction, and specific heat. Nevertheless, the
magnetization expressed in Eq.~(\ref{eq:e8}) indicates that it fails
in exploring magnetic properties. The problem goes more serious in
the case of Bose gases in``synthetic" magnetic field. Neither the
BEC temperature nor the diamagnetization is correctly calculated, as
suggested in  Eqs.~(\ref{eq:e11}) and (\ref{eq:e12}).

The failure of SCA in dealing with magnetic properties of quantum
gases reminds us to recall the Bohr-van Leeuwen theorem. Within this
semi-classical approximation, the grand-canonical free energy of the
ideal Bose gas reads \cite{Kling2007}
\begin{align}
\mathcal{F}=& N_0(\mu_c-\mu) - \sum_{j=1}^{\infty}\frac1{j\beta}
 exp\{-\beta[H_j(\mathbf{x},\mathbf{P})-\mu] \} \nonumber\\
 =&N_0(\mu_c-\mu) - \frac1{\beta} \int \frac{d^3x d^3P} {(2\pi\hbar)^3}
 exp\{ -\beta[H(\mathbf{x},\mathbf{P})-\mu]\},
\end{align}
where $\mu_c$ denotes the critical chemical potential at which the
condensation emerges. The energy levels $H_j$ are replaced by the
classical Hamiltonian
\begin{align}
\hat{H}=\frac{\mathbf{P}^2}{2M}+V(\mathbf{x},\Omega).
\end{align}
where $\mathbf{P}=\mathbf{p}-\mathbf{A}$. Therefore, the gauge
potential $\mathbf{A}$ in the first term of Hamiltonians
(\ref{eq:e1}) and (\ref{eq:e9}) disappears in the integral. In
Hamiltonian (\ref{eq:e1}), the rotation is still partially embodied
in the second term by the rotation frequency $\Omega$, while in
Hamiltonian (\ref{eq:e9}) the rotation information is completely
drowned.

To retrieve the lost information regarding the rotation, the
discrete feature of energy levels must be held back. In the
following section, we will calculate summations over energy levels
directly, rather than converting the summation into integrals.

Recently, EI-Badry has calculated the field dependence of BEC
temperature and the temperature dependence for synthetic
magnetization base on a modified semi-classical approximation. He
obtains the decrease of $\bar{T}_C$ with increasing the field and
$\bar{M}$ with the temperature \cite{EI-Badry2013}. Their results
include simultaneously corrections to quantities arising from the
finite size and interatomic interaction effects. Quantization
feature of energy levels is still not taken into account.

\section {Numerical results: the truncated-summation approach}

We first describe the method employed in this section. Generally,
thermodynamic quantities derived from thermodynamic potential is an
infinite summation over all the energy levels. Take the number of
particles for example,
$N=-{\partial{\bar{\Omega}}}/{\partial{\bar{\mu}}}$, which yields
\begin{align}
N_T=\sum_{nlm}\{\exp[(\bar{\varepsilon}_{nlm}-\bar{\mu})/\bar{T}-1]\}^{-1}
\label{eq:e15}
\end{align}
and
\begin{align}
N_0=-\frac{\partial{\bar{\Omega}_0}}{\partial{\bar{\mu}}}=\left|{\bar{\psi}}\right|^2,
\end{align}
corresponding to the number of thermal and condensed bosons,
respectively. Here $\bar{\varepsilon}_{nlm}$ denotes energy levels
expressed in Eq. (\ref{eq:e2}) for the rotating frame case and Eq.
(\ref{eq:e10}) for ``synthetic" field case, respectively.

Unfortunately, it is impossible to derive an analytical expression
of the infinite summation. Therefore we perform numerical
calculations by truncating the summation at certain order of energy
levels. That is to say, higher energy levels are neglected.
Basically, this truncated-summation approach is reasonable
because the quantum effect we concern occurs at low temperature and
thus the higher energy levels are seldom occupied.

\subsection{Trapped ideal Bose gases in the rotating frame}

Admittedly, the more energy levels are considered, the more accurate
the obtained results are, and the higher the computational cost is.
In order to acquire satisfactory precision of the results, we need
to choose an appropriate cut-off-level of the energy, at which the
summation is truncated. There are three quantum numbers to describe
the energy level, $n$, $l$ and $m$. For simplicity, we set that the
summation over each quantum number is cut off at the same maximum
value $n_{max}$. That is, the scopes of quantum numbers $n, l, m$
are defined as $n\in [0,n_{max}], l\in[0,n_{max}],
m\in[-n_{max},n_{max}]$.

\begin{figure}[htb]
  \centering
  \includegraphics[width=0.45\textwidth]{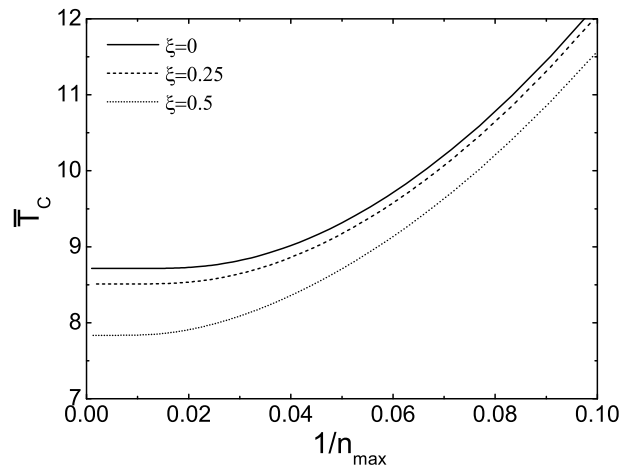}
\caption{Plots of the BEC temperature $\bar{T}_C$ versus the
inverse of the truncated order $n_{max}$ for trapped Bose gases in
the rotating frame.}
 \vspace{-3mm}
\label{fig1}
\end{figure}

Let's first calculate the BEC temperature $\bar{T}_C$ according to
Eq.~(\ref{eq:e15}). The particle number is set as $N=1000$ and
harmonic trap is set to be isotropic ($\sigma=1$). As can be seen
from Fig.~\ref{fig1}, $\bar{T}_C$ drops quickly as $n_{max}$
increases when $n_{max}$ is small, which means that the summation
does not converge. When $n_{max}\geq100$, $\bar{T}_C$ remains stable
with $n_{max}$ increasing. The summation is already convergent
because considering more higher energy levels does not lead to
visible changes to the result. In following calculations, we choose
$n_{max}=1000$ in order to ensure the precision of results, in which
case the total energy levels being considered are up to $10^9$.

\begin{figure}[htb]
  \centering
 \includegraphics[width=0.45\textwidth]{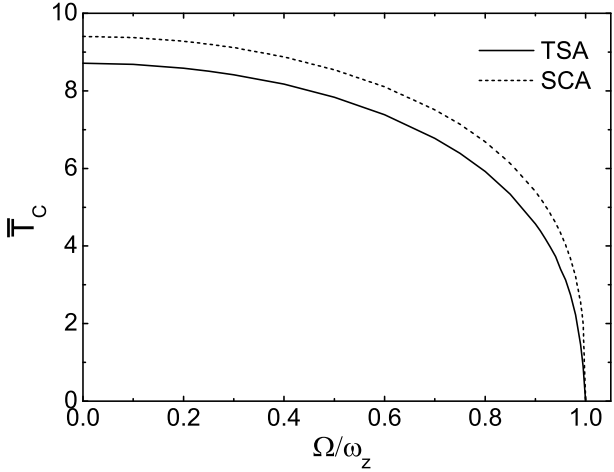}
\caption{The BEC temperature as a function of rotation frequency.
The solid and short dashed lines denote the TSA and SCA results,
respectively.}
 \vspace{-3mm}
\label{fig2}
\end{figure}

We calculated the BEC temperature as a function of the rotation
frequency. The TSA and SCA results are compared in Fig.~\ref{fig2}.
The TSA result is lower than the SCA one, but the difference is
negligible. Particularly, both results drop down to zero when
$\Omega\rightarrow\omega_0$. At this point, the system reaches the
fast-rotating limit and BEC is completely suppressed.

\begin{figure}[htb]
  \centering
 \includegraphics[width=0.45\textwidth]{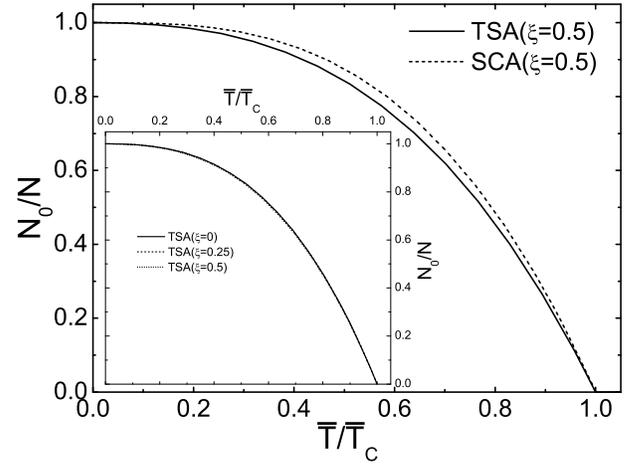}
\caption{Condensate fraction versus the normalized temperature at
$\xi=0.5$. Inset: TSA results for different rotation frequencies.}
 \vspace{-3mm}
\label{fig3}
\end{figure}

The condensate fraction for the $\xi=0.5$ case versus the normalized
temperature is plotted in Fig.~\ref{fig3}. The temperature
dependence of $N_0/N$ deviates slightly from the $1-(\bar{T}/\bar{T}_C)^3$ law
predicted by the SCA. Interestingly, normalized-temperature
dependence of $N_0/N$ at different rotating frequencies obeys almost
the same law, as demonstrated in the inset of Fig.~\ref{fig3}.

\begin{figure}[htb]
  \centering
 \includegraphics[width=0.45\textwidth]{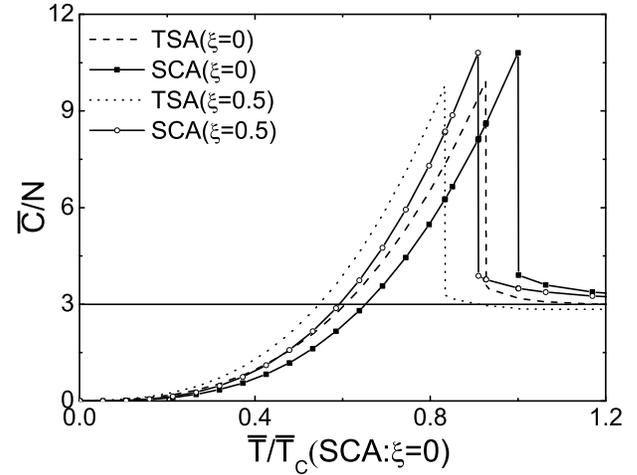}
\caption{The per-particle specific heat versus temperature which is
in unit of the SCA BEC temperature at $\xi=0$. The horizontal solid
line indicates the Dulong-Petit law.}
 \vspace{-3mm}
\label{fig4}
\end{figure}

Figure~\ref{fig4} shows the specific heat, $\bar{C}$,
which is given in the summation form
\begin{align}
\bar{C}=\sum_{nlm}\frac{\bar{\varepsilon}_{nlm}}{\bar{T}^2}\frac{\exp\left(\frac{\bar{\varepsilon}_{nlm}-\bar{\mu}}
{\bar{T}}\right)\left[\bar{T}\frac{\partial\bar{\mu}}{\partial\bar{T}}+\bar{\varepsilon}_{nlm}-\bar{\mu}\right]}
{\left[\exp\left(\frac{\bar{\varepsilon}_{nlm}-\bar{\mu}}{\bar{T}}\right)-1\right]^2}.
\end{align}
For $\bar{T}\leq{\bar{T}_C}$, the chemical potential $\bar{\mu}$ is
equal to the ground state energy $\bar{\varepsilon}_0$. $\bar{C}$
exhibits discontinuity at the critical temperature. It tends to zero
at the low temperature limit $T\rightarrow0$ and Dulong-Petit
specific heat at high temperature region. TSA results are larger
below $\bar{T}_C$, but smaller above $\bar{T}_C$ than the SCA
results.

\begin{figure}[htb]
  \centering
\includegraphics[width=0.45\textwidth]{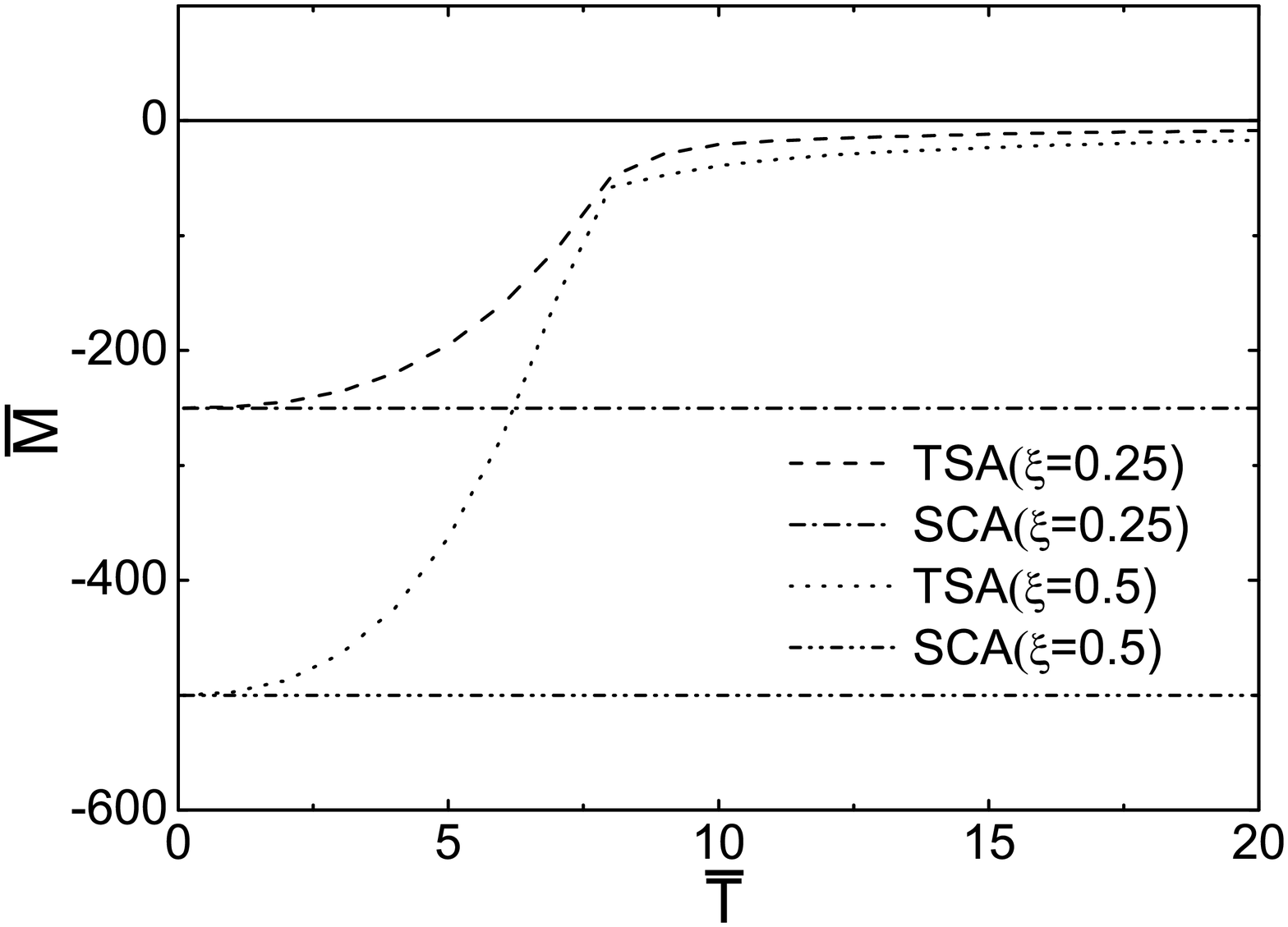}
\caption{Magnetization versus temperature. The solid line indicates
the case of non-rotating states.}
 \vspace{-3mm}
\label{fig5}
\end{figure}

Above results suggest that the TSA only brings about small
corrections to the SCA. We proceed to check the magnetization, which
in the summation form is given by
\begin{align}
\bar{M}=-N_0\xi-
  \sum_{nlm}\frac{(2l+|m|+1)\xi-m}{\exp\left[(\bar{\varepsilon}_{nlm}-\bar{\mu})/\bar{T}\right]-1}.
\end{align}
The obtained result shown in Fig.~\ref{fig5} suggests the TSA
correction to diamagnetization comes dramatically significant.
Distinct from the SCA result which keeps an constant at all
temperatures, $\bar{M}$ becomes varying with $\bar{T}$, particularly
at low temperature. The SCA result just amounts to the minimum value
of the TSA result. $\bar{M}$ drops quickly with $\bar{T}$ dreading
and achieves its maximum at $\bar{T}=0$. This behavior means that
the diamagnetism is stronger at lower temperatures especially below
the transition temperature. Above the BEC temperature, the
$\bar{M}-\bar{T}$ line becomes flatten out, and tends to zero at
higher temperature. This suggests that the rotation effect is sorely
suppressed by thermal fluctuations.

\subsection{Trapped ideal Bose gases in ``synthetic" magnetic field}

In comparison to the gas in the rotating frame, less attention has been
paid to the Bose gas in ``synthetic" magnetic field
\cite{Fan2011,EI-Badry2013}. Plausibly, the two systems might
exhibit similar behaviors, since their Hamiltonians resemble each
other. However, the results based on SCA have revealed that they
show more differences than similarities. Moreover, according to the
analysis in Sec.~\ref{sec1}, the SCA is more unapplicable to the Bose gas
in ``synthetic"  magnetic field. Thus it is in more need of treatment beyond
SCA.

\begin{figure}[htb]
  \centering
  \includegraphics[width=0.45\textwidth]{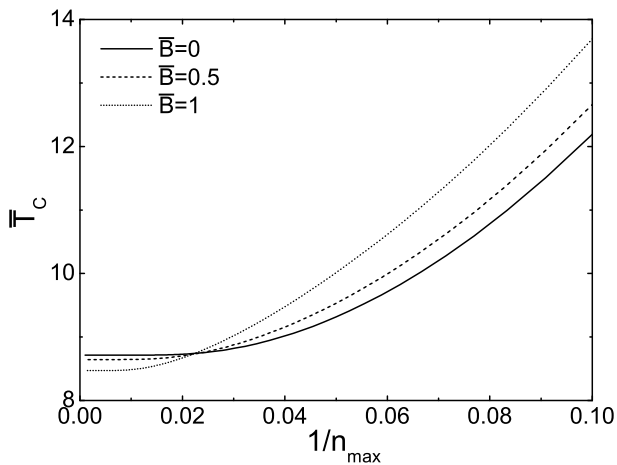}
\caption{The BEC temperature $\bar{T}_C$ versus the
inverse of the truncated order $n_{max}$ at different ``synthetic"
magnetic field $\bar{B}$.}
 \vspace{-3mm}
\label{fig6}
\end{figure}

\begin{figure}[htb]
  \centering
  \includegraphics[width=0.45\textwidth]{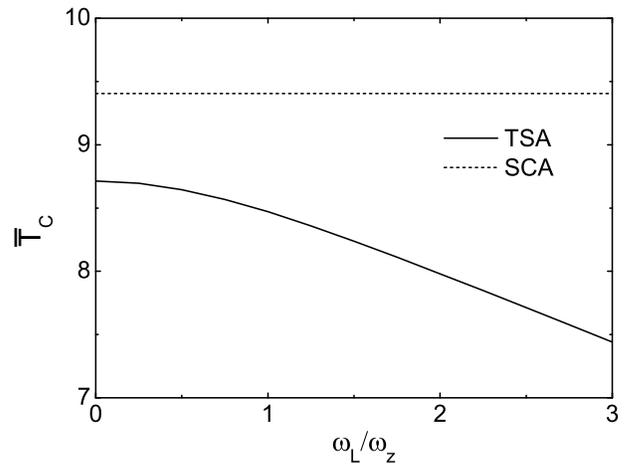}
\caption{The BEC temperatures as a function of
$\omega_L/{\omega_z}$. The solid and short dashed lines show the TSA
and SCA results, respectively.}
 \vspace{-3mm}
\label{fig7}
\end{figure}

\begin{figure}[htb]
  \centering
  \includegraphics[width=0.45\textwidth]{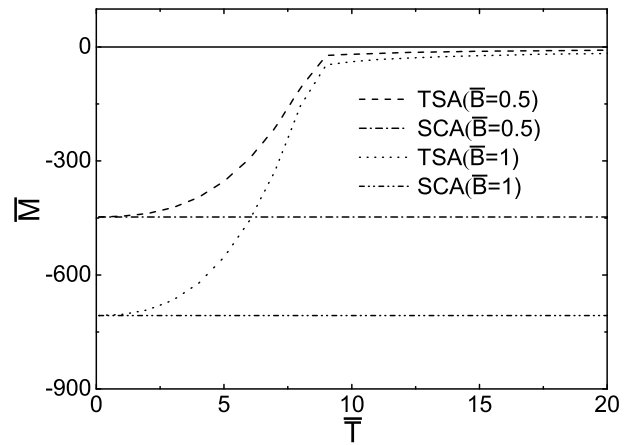}
\caption{Magnetization versus temperature for different ``synthetic"
magnetic field. The solid line denotes the zero-field case.}
 \vspace{-3mm}
\label{fig8}
\end{figure}

Once again, we need to verify the accuracy and reliability of the
TSA. Figure~\ref{fig6} shows the
$\bar{T}_C-1/n_{max}$ curve at different given fields. It should be
noted that convergency of the summation becomes worse in stronger
magnetic field, and thus we need a larger $n_{max}$. And again we
judge that $n_{max}=1000$ is sufficient to ensure acceptable
accuracy.

Figure~\ref{fig7} plots the BEC temperature, $\bar{T}_C$, as a
function of ``synthetic" magnetic field, $\bar{B}$. The ``synthetic"
field-dependence of $\bar{T}_C$ takes on completely different
characteristics from that within the SCA, for which $\bar{T}_C$ has
no relation with $\bar{B}$. The present result appears more
reasonable and keeps consistent in superconductivity when exposed to
a magnetic field, where the transition temperature decreases as the
magnetic field is strengthened. We wish to point out that BEC
temperature of the trapped Bose gas either in the rotating
frame or in ``synthetic" magnetic field decreases with the rotation
frequency $\xi$ or ``synthetic" magnetic field $\bar{B}$.
Nevertheless, $\bar{T}_C$ drops more quickly in the former case, as
indicated in Figs.~\ref{fig2} and \ref{fig7}.

As shown in Fig.~\ref{fig7}, $\bar{T}_C\approx 7.4$ at $\bar{B}=3$,
which is still quite large. Based on present results, we suspect
that $\bar{T}_C\to 0$ at about $\bar{B}=18$ according to the
decreasing trend of the BEC temperature with ``synthetic"
magnetic field. However, within the current truncated energy levels,
we can not find $\bar{T}_C\to 0$. This might be owing to the fact
that $n_{max}$ is not large enough to produce convincing numerical
results at that large field. If stronger field $\bar{B}$ is
considered, we need to choose a larger $n_{max}$.

Figure~\ref{fig8} shows the ``synthetic" diamagnetism calculated
from the following formula,
\begin{align}
\bar{M}=-\frac{N_0\bar{B}}{\sqrt{\sigma^2+\bar{B}^2}}
  -\sum_{nlm}\frac{(2l+|m|+1) \frac{\bar{B}}{\sqrt{\sigma^2+\bar{B}^2}}-m}
    {\exp\left[(\bar{\varepsilon}_{nlm}-\bar{\mu})/\bar{T}\right]-1}.
\end{align}
Its feature is similar to that of the trapped Bose gas in the
rotating frame. It is by no means a constant with temperature as the
SCA predicted. The ``synthetic" magnetization $\bar{M}$ decreases as
the temperature is lowered, reaching its maximum value at
$\bar{T}=0$. Magnetization $\bar{M}$ is stronger and negative at
larger ``synthetic" magnetic field $\bar{B}$ below the BEC
temperature. Our results provide a theoretical support for the
current experiment \cite{Pasquiou2012}.

\subsection{Comparison with other improvements to the semi-classical approximation}

It is worth noting that two other revelent approaches have already
been proposed to improve the SCA by taking into account, more or
less explicitly, the quantization feature of energy levels
\cite{Pelster2010,Pelster2009}. In Ref. \cite{Pelster2010},
Bala\u{z} {\it et al.} developed an efficient ultra-fast converging
path-integral approach on the basis of precise single-particle
energy spectrum and obtained more reasonable results than
semi-classical calculations, in particular for smaller particle
numbers. The other approach retrieves the lost information in SCA by
considering higher order corrections when replacing summations by an
integral \cite{Pelster2009}. Thermodynamic properties of
harmonically trapped ideal Bose gases without rotation were
calculated analytically.

\begin{figure}[htb]
  \centering
  \includegraphics[width=0.45\textwidth]{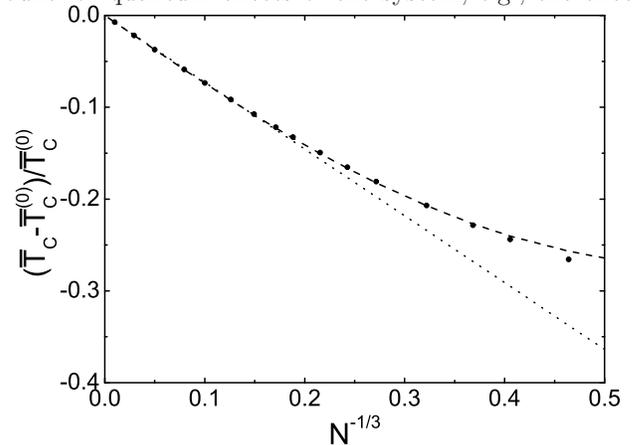}
\caption{Finite-size corrections
$(\bar{T}_C-\bar{T}_C^{(0)})/\bar{T}_C^{(0)}$ versus $N^{-1/3}$. The
bullets denote the TSA correction. $\bar{T}_C$ and $\bar{T}_C^{(0)}$
denote the BEC temperature for ideal gases without rotation within
the TSA and SCA, respectively. The dotted (dashed) line corresponds
to the first (the first two) finite-size corrections which is taken
from Eq. (32) in Ref.~\cite{Pelster2009}.}
 \vspace{-3mm}
\label{fig9}
\end{figure}

Relatively, the TSA is a quite simple but already effective approach
to go beyond the SCA. In order to demonstrate this point, we give a
direct comparison between the TSA result and the first (the first
two) finite-size corrections in Ref. \cite{Pelster2009}, as shown in
Fig.~\ref{fig9}. It is clearly shown that the TSA correction yields
very good agreement with the first two finite-size correction even
for small particle numbers.

\section {Conclusions}

In this paper, we study thermodynamic properties of the rotating
trapped ideal Bose gas numerically using a truncated-summation
approach. This approach conserves the discrete feature of energy
levels in contrast to the semi-classical approximation, thus it helps
to account for quantum effects of the system, e.g., the effective
``magnetization" resulted from the rotation. The obtained results
indicate that the ``diamagnetization" turns significantly stronger
below the BEC temperature than above, coinciding with our intuition
established in superconductivity.

We perform calculations of the BEC temperature as a function of the
rotation frequency for bosons spined up both by the rotating frame
and ``synthetic" magnetic field. In both cases, the BEC temperature
decreases with the increasing rotation frequency or ``synthetic"
magnetic field, which suggests that the rotation tends to destroy
BEC. The decrease seems much slower in the latter case. It probably
implies that more stronger ``synthetic" magnetic field is needed in
order to drive the system to the fast-rotating limit. Moreover, the
specific heat is briefly discussed. Similar to the SCA result, the
specific heat shows discontinuity at the BEC temperature and tends
to a constant in high temperature region.

\section*{Acknowledgements}

This work is supported by the National Key Basic Research Program of
China (Grant No. 2013CB922000), the National Natural Science
Foundation of China (Grant No. 11074021) and the Fundamental
Research Funds for the Central Universities of China. QG
acknowledges support from the China Scholarship Council and the
hospitality of the Rudolf Peierls Centre for Theoretical Physics,
University of Oxford.

\end{document}